\documentclass[conference]{IEEEtran}
\IEEEoverridecommandlockouts
\usepackage{cite}
\usepackage{amsmath,amssymb,amsfonts}

\usepackage{algorithmic}
\usepackage{graphicx}
\usepackage[normalem]{ulem}
\usepackage{textcomp}
\usepackage{xcolor}
\usepackage{physics}
\usepackage{float} 
\usepackage{caption}
\usepackage{placeins}
\usepackage{subcaption}
\usepackage{tikz}
\usepackage{adjustbox}
\usetikzlibrary{3d,calc}
\usepackage{tikz-3dplot}
\def\BibTeX{{\rm B\kern-.05em{\sc i\kern-.025em b}\kern-.08em
    T\kern-.1667em\lower.7ex\hbox{E}\kern-.125emX}}
\begin{document}

\title{The Better Solution Probability Metric: Optimizing QAOA to Outperform its Warm-Start Solution\\
}

\newcommand{\rt}[1]{{\color{blue}[(Reuben) #1}]}
\newcommand{\rte}[1]{{\color{blue} #1}}
\newcommand{\cut}{\text{cut}}

\definecolor{kellygreen}{rgb}{0.3, 0.73, 0.09}

\newcommand{\se}[1]{{\color{kellygreen}[(Stephan) #1}]}
\newcommand{\see}[1]{{\color{kellygreen} #1}}

\newcommand{\sfe}[1]{{\color{red}[(Sean) #1}]}
\newcommand{\sfee}[1]{{\color{red} #1}}

\author{
\IEEEauthorblockN{Sean Feeney}
\IEEEauthorblockA{\textit{CCS-3 Information Sciences} \\
\textit{Los Alamos National Laboratory}\\
Los Alamos, NM, 87544 USA \\
sfeeney@lanl.gov}
\and

\IEEEauthorblockN{Reuben Tate}
\IEEEauthorblockA{\textit{CCS-3 Information Sciences} \\
\textit{Los Alamos National Laboratory}\\
Los Alamos, NM, 87544 USA \\
rtate@lanl.gov}
\and
\IEEEauthorblockN{Stephan Eidenbenz}
\IEEEauthorblockA{\textit{CCS-3 Information Sciences} \\
\textit{Los Alamos National Laboratory}\\
Los Alamos, NM, 87544 USA \\
eidenben@lanl.gov}

}

\maketitle

\begin{abstract}
This paper presents a numerical simulation investigation of the Warm-Start Quantum Approximate Optimization Algorithm (QAOA) as proposed by Tate et al. \cite{tate2024guaranteeswarmstartedqaoasingleround}, focusing on its application to 3-regular Max-Cut problems. Our study demonstrates that Warm-Start QAOA consistently outperforms theoretical lower bounds on approximation ratios across various tilt angles, highlighting its potential in practical scenarios beyond worst-case predictions. Despite these improvements, Warm-Start QAOA with traditional parameters optimized for expectation value does not exceed the performance of the initial classical solution.
To address this, we introduce an alternative parameter optimization objective, the Better Solution Probability (BSP) metric. Our results show that BSP-optimized Warm-Start QAOA identifies solutions at non-trivial tilt angles that are better than even the best classically found warm-start solutions with non-vanishing probabilities.  
These findings underscore the importance of both theoretical and empirical analyses in refining QAOA and exploring its potential for quantum advantage.
\end{abstract}

\begin{IEEEkeywords}
Warm-Start, QAOA, Max-Cut
\end{IEEEkeywords}

\section{Introduction}
The forefront of the field of quantum algorithms is focused on the achievement of quantum advantage with modern quantum devices\cite{bharti2022noisy},\cite{keller2024quantumapproximateoptimizationcomputational},\cite{lau2022nisq}. Within the past decade, there have been demonstrations that claim quantum advantage or supremacy, e.g., by generating distributions obtained by randomized quantum circuits; since then, there has been a shift towards demonstrating quantum advantage for more practical problems. Particular problems of interest that could provide significant value to both scientific advancements and commercialization of the field as a whole are optimization problems \cite{abbas2023quantumoptimizationpotentialchallenges}. This paper focuses on a sub division of problems known as combinatorial optimizaiton problems; this class of problems includes famous examples such as the Traveling Salesman Problem, Max-Cut, Minimum Spanning Tree, among many others. These problems have applications in logistics \cite{BLN18}, VLSI \cite{barahona1988application}, power systems \cite{soares2018survey}, and beyond. Of particular interest in this paper are quantum algorithms with provable approximation ratio guarantees on combinatorial optimization problems such as Max-Cut. One such algorithm proposed for solving these types of combinatorial optimization problems is the Quantum Approximation Optimization Algorithm, (QAOA), put forward by Farhi et al. in 2014\cite{farhi2014quantumapproximateoptimizationalgorithm}. It is part of the hybrid classical-quantum variational algorithms schemes that have been developed and researched heavily upon over the last decade. \cite{cerezo2021variational}

Theoretical guarantees for QAOA are often difficult to come by; current results  only provide results for extremely low-depths and/or specific graph families (e.g. Farhi's 0.6924-approximation for 3-regular graphs at circuit depth $p=1$ \cite{farhi2014quantumapproximateoptimizationalgorithm}). Meanwhile, theoretical negative results \cite{farhi2020quantum, farhi2020quantum2,bravyi2020obstacles, hastings2019classical,marwaha2021local} suggest that the QAOA algorithm may need to be modified if one hopes to observe any kind of quantum advantage, especially in the current regime of quantum devices where deep circuit depths are infeasible to work with due to the compounding effects of quantum noise.


One promising modification of QAOA is known as Warm-Start QAOA where the initial quantum state is modified from $\ket{+}^{\otimes n}$ (the all-equal superposition of bitstrings) to an alternate starting state, typically one that is designed to be biased towards better solutions. While various Warm-Start approaches have been considered \cite{egger2021warm, 10.1145/3549554, Tate_2023, tate2024guaranteeswarmstartedqaoasingleround,cain2022qaoa}, this work focuses on the approach that Tate et al. \cite{tate2024guaranteeswarmstartedqaoasingleround} developed in 2024, which is the first Warm-Start approach with known theoretical worst-case guarantees at depth $p=1$. Given a tilt angle parameter $\theta$ and a bitstring $b$ (which corresponds to some cut) that is locally optimal up to single bitflips, Tate et al. \cite{tate2024guaranteeswarmstartedqaoasingleround} construct an initial warm-start state $\ket{b_\theta}$ which becomes increasingly biased towards $\ket{b}$ as $\theta \to 0$; the approach is equivalent to standard QAOA at $\theta = 90^\circ$. Tate et al. \cite{tate2024guaranteeswarmstartedqaoasingleround} also alter the mixing Hamiltonian in a way that allows QAOA to still be viewed as a Trotterization of Quantum Adiabatic Computing. For various values of $\theta$, Tate et al. \cite{tate2024guaranteeswarmstartedqaoasingleround} numerically calculate theoretical lower bounds on the worst-case approximation ratios for single-round ($p=1$) Warm-Start QAOA on 3-regular Max-Cut instances.

Overall, the theoretical results by Tate et al. \cite{tate2024guaranteeswarmstartedqaoasingleround} paint a grim picture for Warm-Start QAOA. However, it is often the situation that worst-case guarantees are not representative of the behavior of algorithms on more ``typical" instances; for example, the simplex method for solving problems in linear programming is known to run in exponential time in the worst case but is incredibly efficient in practice \cite{spielman2004smoothed}. Thus, it may still be possible that Warm-Start QAOA has a \emph{practical} advantage; this motivates our work which serves as an experimental exploration of Tate et al.'s \cite{tate2024guaranteeswarmstartedqaoasingleround}  warm-start approach.


In this work we contribute three main findings: 
\begin{enumerate}
    \item First, we find that single-round Warm-Start QAOA on randomly generated 3-regular instances \emph{significantly} outperforms the theoretical lower bounds obtained by Tate et al. \cite{tate2024guaranteeswarmstartedqaoasingleround}, with the empirical results consistently achieving a higher averaged approximation ratio for any choice of tilt angle $\theta$. 
    \item Second, we show that, regardless of the parameter-finding scheme used, for all graphs $G$ and for all initial bitstrings $b$ tested, there does not exist a tilt angle $\theta \in \{1^\circ, 2^\circ, \dots, 90^\circ\}$ for which Warm-Start QAOA produces an approximation ratio greater than what is achieved by the initial classical cut corresponding to $b$.
    \item Lastly, we propose a novel figure-of-merit which we call \emph{Better Solution Probability} (BSP). For BSP, we simply change the objective function of  the QAOA parameter finding step from maximizing expectation value to maximizing the probability of finding a solution that is better than the initially provided warm-start.  We empirically show that, with the BSP metric, Warm-Start QAOA frequently performs best at non-trivial tilt angles $0^\circ < \theta < 90^\circ$, beating both the classical cut ($\theta = 0^\circ$) and standard QAOA ($\theta = 90^\circ$). BSP-QAOA reliably finds better solutions with a small, but non-vanishing probability. Moreover, we show that the BSP metric is practical in the sense that it is possible to optimize the circuit parameters for this metric (unlike other metrics such as ground state probability which requires knowledge of the optimal solution ahead of time).
\end{enumerate}

\section{Related Work}
Various warm-start techniques for QAOA have been introduced. The first Warm-Start QAOA variants were proposed by Egger et al.\cite{egger2021warm} which refer to as \emph{Continuous Warm-start QAOA} and \emph{Rounded Warm-Start QAOA}. Their Continuous Warm-Start QAOA uses the relaxation of a QUBO (Quadratic Unconstrained Binary Optimization); however, it should be noted that their QUBO relaxation is only convex (i.e. easily solvable) under certain circumstances (which do not hold for Max-Cut). Their Rounded Warm-Start QAOA approach is similar to the approach in this work in that a \emph{particular} solution and a tilt angle $\theta$ is used to create the initial warm-start quantum state; however, they instead primarily focused on an unconventional unaligned mixer. At $\theta=\pi/3$, for Max-Cut, this unaligned mixer (with suitable circuit parameters) is able to recover the classical solution used to create the warm-start; however, unlike aligned mixers (defined in Section \ref{sec:warmStartQAOA}), this approach has no convergence guarantees as $p\to \infty$.

Independently from Egger et al. \cite{egger2021warm}, Tate et al. had developed an alternative warms-start technique based on Burer-Monteiro (BM) \cite{burer2003nonlinear} and Goemans-Williamson (GW) \cite{GW95} relaxations  of the Max-Cut problem; they considered the Pauli-$X$ mixer in their first paper \cite{10.1145/3549554} and improved aligned mixers in a follow-up paper \cite{Tate_2023}. While such GW-based warm-starts yielded good empirical results, they found that such a warm-start technique was difficult to theoretically analyze beyond $p=0$ (i.e. simply measuring the warm-start state).

Recent work by Tate et al. \cite{tate2024guaranteeswarmstartedqaoasingleround} considered warm-starts similar to Egger et al.'s Rounded Warm-Start QAOA approach but with aligned mixers and with alternative methods for obtaining a classical solution (to later use to construct the warm-start) on 3-regular Max-Cut instances. As stated earlier in the introduction, this line of work \cite{tate2024guaranteeswarmstartedqaoasingleround} was primarily theoretical in nature and this work serves as a experimental exploration of this specific Warm-Start QAOA approach; we provide more details regarding this recent work by Tate et al. \cite{tate2024guaranteeswarmstartedqaoasingleround} in Section \ref{sec:warmStartQAOA}.

Lastly, we note that our BSP novel metric, discussed earlier in the introduction, bears some similarity to the Threshold-Based QAOA \cite{Golden_2021} introduced by Golden et al. which considers solutions above some specificed threshold value; we discuss the differences between QAOA with BSP optimization and Threshold-Based QAOA in more detail in Section \ref{sec:novel_metric}.
\section{Background}

\subsection{QAOA}

The Quantum Approximation Optimization Algorithm (QAOA) was developed as a method to provide approximate solutions to combinatorial optimization problems \cite{farhi2014quantumapproximateoptimizationalgorithm}. Combinatorial optimization problems are frequently given over binary inputs $x\in \{0,1\}^{n}$, where $f:\{0,1\}^{n} \to \mathbb{R}$ is an objective function that evaluates the cost of solution $x$. The goal is to maximize or minimize the function $f(x)$ to achieve an approximate solution and prepare a state $\ket{\psi}$ that one can sample high quality solutions from. The QAOA is comprised of four main components:

\begin{itemize}
    \item A cost Hamiltonian $H_{c}$, diagonal in the computational basis: $H_{c}\ket{x}=f(x)\ket{x}$, referred to as a phase separator 
    \item A mixing Hamiltonian $H_{m}$
    \item An integer $p \geq 1$ to represent the number of layers in the QAOA circuit
    \item Two real arrays of circuit parameters\footnote{Much of the QAOA literature refer to the $\gamma$ and $\beta$ parameters as ``angles"; however, we avoid this terminology in order to avoid potential confusion with the tilt angle $\theta$ used in Warm-Start QAOA.} $\beta = \{ \beta_{1},\dots \beta_{p}\}$, $\gamma = \{ \gamma_{1},\dots \gamma_{p}\}$, of which the application of both $H_{c}$ and $H_{m}$ are dependent upon
    \item An initial state $\ket{s_{0}}$
\end{itemize}
In the context of standard Max-Cut QAOA with graph $G=(V,E)$ and edge weights $w: E \to \mathbb{R}$, the Hamiltonians and the initial state are given by:
\begin{equation}
    H_{c} = \frac{1}{2}\sum_{(i,j) \in E} w_{ij}(-\sigma^{z}_{j}\sigma^{z}_{k}+1),
\end{equation}
\begin{equation}
    H_{m} = \sum^{n}_{i} \sigma^{x}_{i},
    \label{eq:pauli_x}
\end{equation}

\begin{equation}
    \ket{s_{0}} = \ket{+}^{\otimes n} = \frac{1}{2^{n}} \sum_{z} \ket{z},
\end{equation}
where $\sigma^{x}, \sigma^{y}, \sigma^{z}$ are the standard Pauli operators. Equation \ref{eq:pauli_x} is often referred to as the Pauli-$X$ mixer.
The initial state $\ket{s_{0}}$ is the uniform superposition over the computational basis states measured in the $z$ basis. In the context of Max-Cut, all $2^n$ possible cuts have the same probability of being observed when measuring $\ket{s_0}=\ket{+}^{\otimes n}$ in the $z$ basis.

The QAOA circuit is constructed by starting with $\ket{s_0}$ and iteratively applying unitaries corresponding to $H_B$ and $H_C$ in an alternating fashion for $p$ rounds:
\begin{equation}
   \ket{\gamma,\beta}=e^{-i\beta_{p} H_{m}}e^{-i\gamma_{p} H_{c}}\dots e^{-i\beta_{1} H_{m}}e^{-i\gamma_{1} H_{c}} \ket{s_{0}}. 
\end{equation}
Each cost unitary $e^{-i\gamma_j H_C}$ applies a phase to each basis state $\ket{x}$ (with $x\in \{0,1\}^n$) that is proportional to the classical objective value $f(x)$. Each mixing unitary $e^{-i\beta_j H_B}$ applies a single-qubit rotation about the $x$-axis of the Bloch sphere to every qubit (with the rotation angle determined by $\beta_j$).
QAOA, as a hybrid quantum classical routine, relies on a classical optimizer to tune the circuit parameters $\beta, \gamma$ maximizing the expectation value $\bra{\gamma,\beta}H_c\ket{\gamma,\beta}$.

\subsection{Classical Max-Cut}
\label{sec:classicalMaxCut}
The Max-Cut problem is as follows: given a graph $G=(V,E)$ and edge weights $w: E \to \mathbb{R}$, partition the vertex set $V$ into two disjoint groups so that the sum of the weights of the edges between the groups is maximized. If $|V|=n$, a feasible solution can be represented as a bitstring $b\in \{0,1\}^n$ where the value of $j$th bit corresponds to one of the two possible parts of the partition. Algebraically, the Max-Cut problem is equivalent to the following maximization problem:
$$\text{Max-Cut}(G) = \max_{b\in \{0,1\}^n} \text{cut}(b),$$
$$\text{cut}(b) = \frac{1}{2}\sum_{(i,j) \in E} w_{ij} \cdot \mathbf{1}[b_i \neq b_j].$$
We say that a bitstring $b$ is locally optimal\footnote{We will later also consider local \emph{continuous} optimization of the circuit parameters. While it should be clear from context what is meant, the reader should be careful not to conflate these concepts with one another.} (LO) if $\text{cut}(b) \geq \text{cut}(b^{(j)})$ for all $j\in [n]$, where $b^{(j)}$ denotes the bitstring $b$ with the $j$th bit flipped; such cuts have the property that moving a single vertex to the other side of the cut cannot improve the cut value. For unweighted 3-regular graphs, the theoretical lower bounds on the approximation ratio by Tate et al. \cite{tate2024guaranteeswarmstartedqaoasingleround} assume that the initial cut $b$ for Warm-Start QAOA is an LO cut. In this work, we obtain LO cuts by starting with a random bitstring (chosen uniformly at random) and iteratively flipping bits to improve the cut value until it is no longer possible to do. It is straightforward to show that for (unweighted) 3-regular graphs, algorithms that produce LO cuts have an approximation ratio of $2/3$, see \cite{tate2024guaranteeswarmstartedqaoasingleround} for a proof.

The best known algorithm with provable approximation-ratio guarantees is the Goemans-Williamson (GW) algorithm \cite{GW95} which is a randomized classical algorithm with a worst-case approximation ratio of 0.878 (in expectation) for graphs with non-negative edge weights. The GW algorithm works by solving a semidefinite program (SDP) relaxation and then performing a randomized hyperplane-rounding procedure on the SDP solution.

Recall that the theoretical lower bounds by Tate et al. \cite{tate2024guaranteeswarmstartedqaoasingleround} require that the initial cut used is a LO cut; however, cuts obtained by the GW algorithm may not necessarily be LO cuts. For this reason, for any cut obtained by the GW algorithm, we perform the same local search described earlier (using the cut returned by the GW algorithm as a starting point) until we obtain a locally optimal cut. We will use the term \emph{GW local search cuts} to refer to cuts obtained by this overall procedure.



\subsection{Approximation Ratio}
Given a bitstring $b$ corresponding to a cut in a graph $G$, the approximation ratio associated with $G$ and $b$ is defined as $\text{cut}(b)/\text{Max-Cut}(G)$. Given a graph $G$ and some output state $\ket{\psi}$ of some quantum circuit, we define the approximation ratio associated with $G$ and $\ket{\psi}$ as $\bra{\psi}H_C\ket{\psi}/\text{Max-Cut}(G)$ where $H_C$ is the Max-Cut cost hamiltonian for $G$.

\subsection{Warm-Start QAOA}
\label{sec:warmStartQAOA}
As in Tate et al. \cite{tate2024guaranteeswarmstartedqaoasingleround}, we consider warm-starts of the form $\ket{s_0} = \ket{b_\theta}$ where $b$ is a classically obtained bitstring and $\theta \in [0, \pi]$ is a parameter referred to as the tilt angle. In the context of Max-Cut QAOA, we refer to $b$ as the initial (warm-start) cut. Geometrically, $\ket{b_\theta}$ is a product-state where the $j$th qubit is at an angle $\theta$ away from either the north or south pole of the Bloch sphere depending on the value of $b_j$ (the $j$th bit of $b$).

Additionally, just as in \cite{tate2024guaranteeswarmstartedqaoasingleround}, we adapt the QAOA mixing Hamiltonian $H_m$ so that it is \emph{aligned} with the warm-start $\ket{b_\theta}$, i.e., $\ket{b_\theta}$ is a ground state of $H_m$; this ensures (due to QAOA's connection to the adiabatic theorem \cite{farhi2014quantumapproximateoptimizationalgorithm}) that as $p \to \infty$, that an optimal cut is sampled with probability approaching 1. Geometrically, these aligned mixers correspond to single-qubit rotations about each qubit's original Bloch sphere position in $\ket{b_\theta}$.

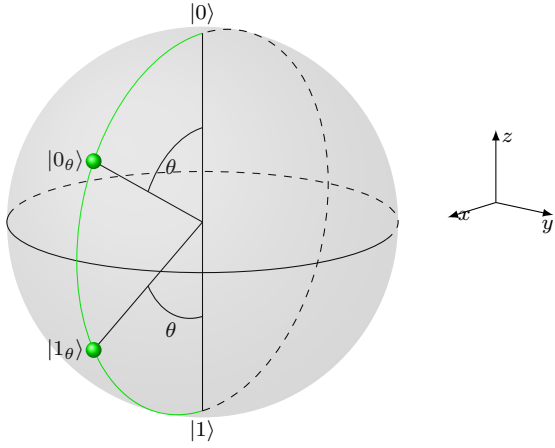
\begin{figure}
\centering
    \pgfmathsetmacro{\r}{2.6} %

\tdplotsetmaincoords{75}{130}
\begin{tikzpicture}[
tdplot_main_coords,
font=\footnotesize,
Helpcircle/.style={gray!70!black,
},
]

\pgfmathsetmacro{\h}{0.9*\r} %

\pgfmathsetmacro{\t}{60}
\coordinate[label=left:{$\ket{0_\theta}$}] (X1) at ({\r*sin(\t)},0,{\r*cos(\t)});
\coordinate[label=left:{$\ket{1_\theta}$}] (X2) at ({\r*sin(\t)},0,{-\r*cos(\t)}); 

\coordinate (M) at (0,0,0);
\coordinate[label=$\ket{0}$] (Top) at (0,0,\r);
\coordinate[label=below:{$\ket{1}$}] (Bot) at (0,0,-\r);

\tdplotdrawarc{(M)}{\r}{-65}{110}{anchor=north}{}
\tdplotdrawarc[dashed]{(M)}{\r}{110}{295}{anchor=north}{}


\tdplotsetrotatedcoords{90}{90}{0}%
\tdplotdrawarc[tdplot_rotated_coords, green]{(M)}{\r}{180}{360}{anchor=north}{}
\tdplotdrawarc[tdplot_rotated_coords, dashed]{(M)}{\r}{0}{180}{anchor=north}{}
\tdplotdrawarc[tdplot_rotated_coords]{(M)}{0.5*\r}{180}{180+\t}{anchor=north}{$\theta$}
\tdplotdrawarc[tdplot_rotated_coords]{(M)}{0.5*\r}{0}{-\t}{anchor=north}{$\theta$}
\draw[] (M) -- (X1);
\draw[] (M) -- (X2);
\draw[] (Top) -- (Bot);

\begin{scope}[tdplot_screen_coords]
\fill[ball color= gray!20, opacity = 0.1] (M) circle (\r); 
\end{scope}

\foreach \P in {X1,X2}{
\shade[ball color=green] (\P) circle (3pt);
}

\begin{scope}[-latex, shift={(M)}, xshift=1.5*\r cm, yshift=0.1*\r cm]
\foreach \P/\s/\Pos in {(1,0,0)/x/right, (0,1,0)/y/below, (0,0,1)/z/right} 
\draw[] (0,0,0) -- \P node[\Pos, pos=0.9,inner sep=2pt]{$\s$};
\end{scope}

\end{tikzpicture}
\caption{The states $\ket{0_\theta}$ and $\ket{1_\theta}$ geometrically depicted on the Bloch sphere. The green half-circle, $\textbf{Arc}$, in the $xz$-plane denotes all the possible positions for $\ket{0_\theta}$ and $\ket{1_\theta}$ as $\theta$ varies from $0$ to $\pi$. }
\label{fig:warmstartVisualization}
\end{figure}

A visualization of the warm-start state $\ket{b_\theta}$ and the action of the corresponding aligned mixer $H_m$ can be seen in Figure \ref{fig:warmstartVisualization}. We refer the reader to Tate et al. \cite{tate2024guaranteeswarmstartedqaoasingleround} for further details regarding the construction and implementation of the warm-start state and mixer.

For various values of $\theta \in [0,\pi]$, Tate et al. \cite{tate2024guaranteeswarmstartedqaoasingleround} determine lower bounds on the worst-case approximation ratio for single-round Warm-Start QAOA on 3-regular graphs under the assumption that the initial cut is an LO cut. More specifically, for each $\theta$, they numerically determine the optimal ($\theta$-dependent) choice of $\gamma,\beta$ so that the bound on the worst-case approximation ratio (with a fixed choice of $\gamma$ and $\beta$) is as best as possible. The values of these lower bounds and optimal circuit parameters can be found in Table 2 in Tate et al.'s work \cite{tate2024guaranteeswarmstartedqaoasingleround}. We will refer to these bounds and angles as \emph{Tate's theoretical lower bounds} and \emph{Tate's theoretical circuit parameters} respectively.

\section{Experimental Setup}
\label{sec:experimentalSetup}

\begin{figure*}[t]
  \centering
  \subfloat[Average approximation ratio for basin-hopping circuit parameter optimization with \textbf{Local Search} warm-start state.]{%
    \includegraphics[width=0.49\textwidth]{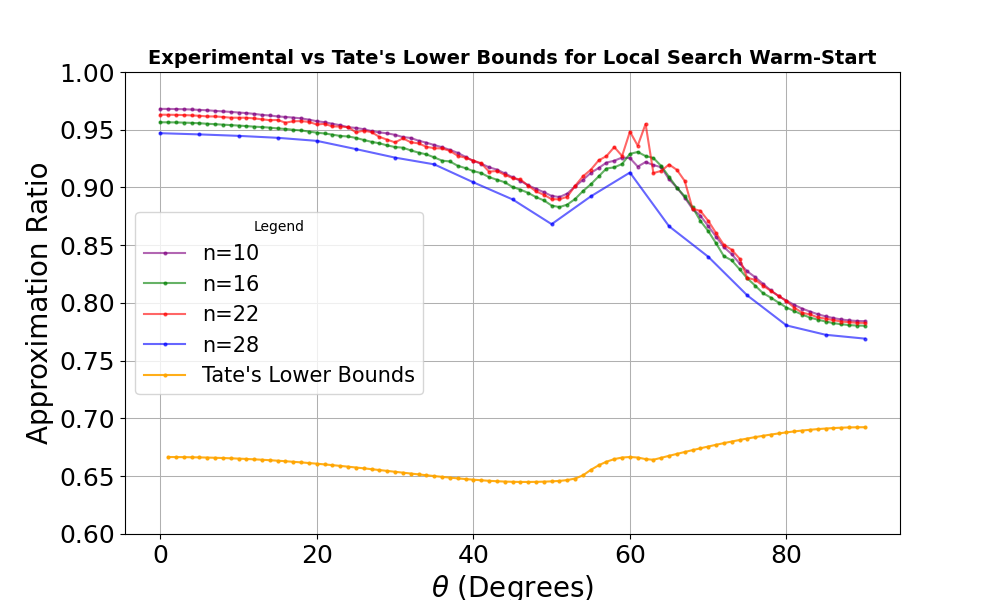} 
  }%
  \hfill
  \subfloat[Average approximation ratio for basin-hopping circuit parameter optimization with \textbf{Goemans-Williamson Local Search} warm-start state.]{%
    \includegraphics[width=0.49\textwidth]{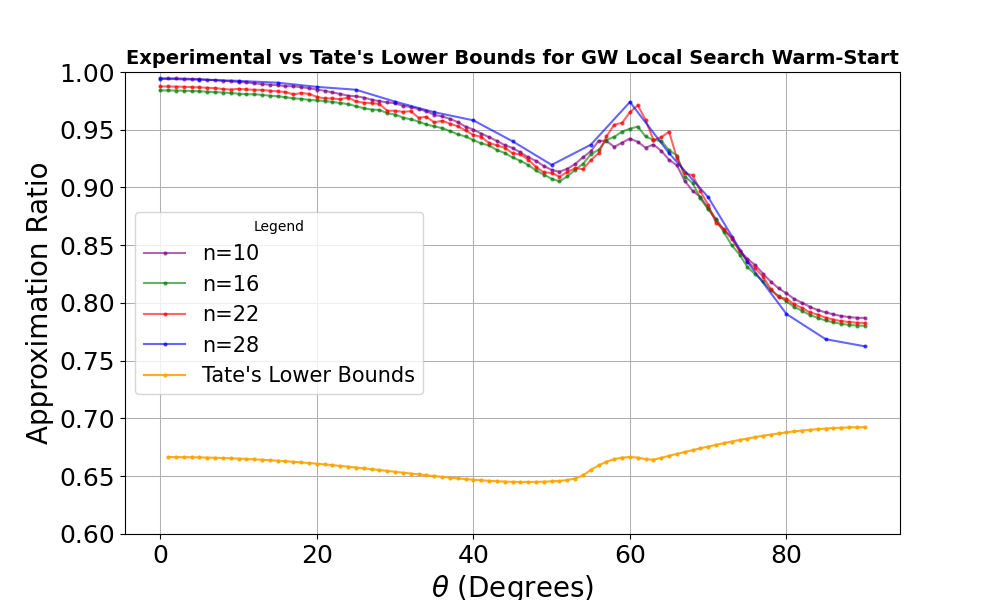}%
  }\\
  \vspace{0.5cm} 
  \subfloat[Average approximation ratio vs problem size $n$ basin-hopping circuit parameter optimization with \textbf{Local Search} warm-start state for $\theta = 0^\circ , 45^\circ, 60^\circ, 90^\circ$ tilt angles. Red dashed lines represent the best known 3-Regular Max-Cut approximation ratio \cite{halperin2004max} and Farhi's QAOA 3-Regular Max-Cut approximation ratio guarantee for circuit depth $p=1$. ]{%
    \includegraphics[width=0.49\textwidth]{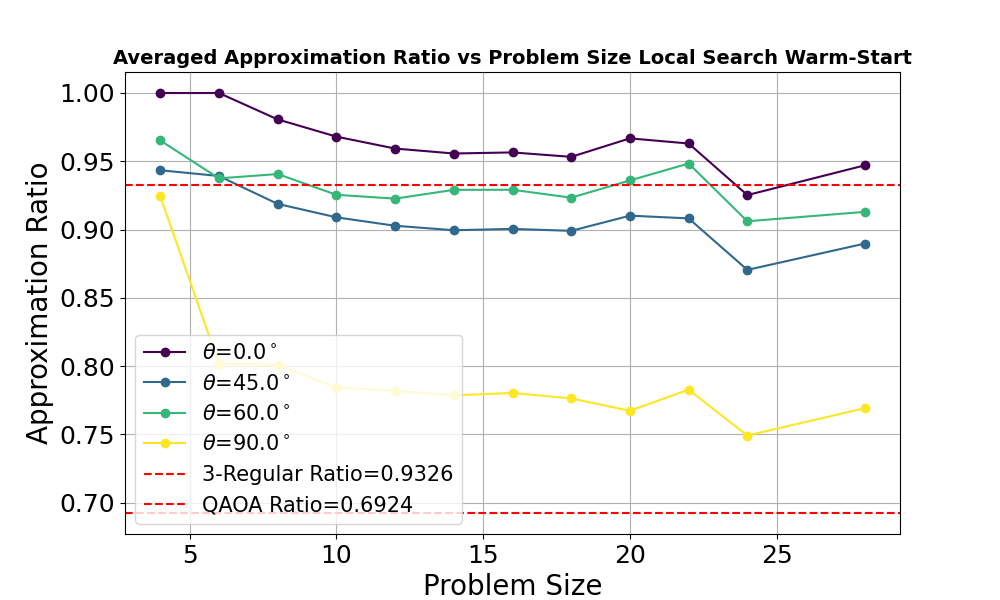}%
  }%
  \hfill
  \subfloat[Average approximation ratio vs problem size $n$ basin-hopping circuit parameter optimization with \textbf{Goemans-Williamson Local Search} warm-start state for $\theta = 0^\circ , 45^\circ, 60^\circ, 90^\circ$ tilt angles. Red dashed lines represent the best known 3-Regular Max-Cut approximation ratio \cite{halperin2004max} and Farhi's QAOA 3-Regular Max-Cut approximation ratio guarantee for circuit depth $p=1$. ]{%
    \includegraphics[width=0.49\textwidth]{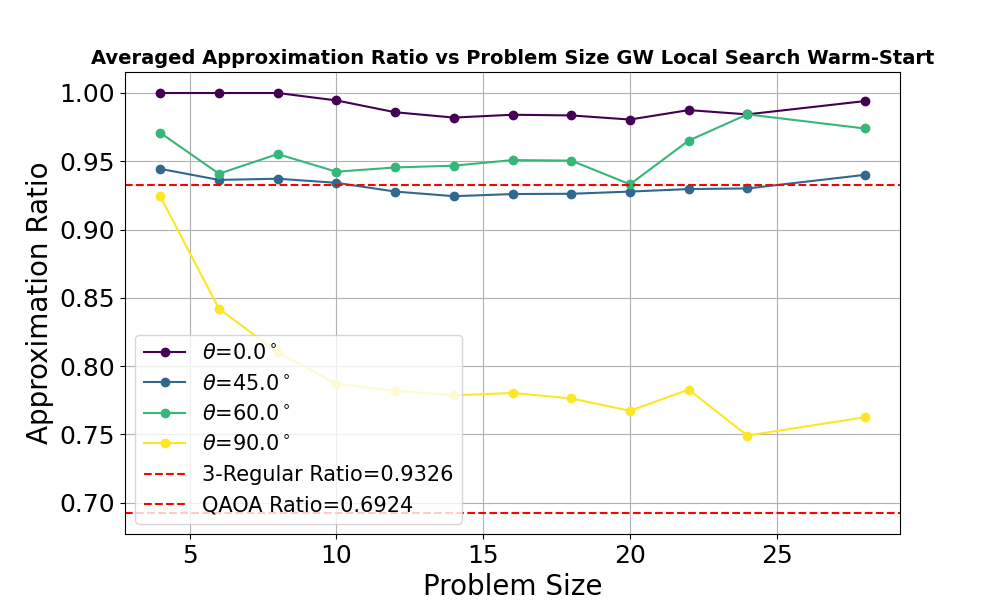}%
  }
  
  \caption{}
  \label{fig:bh_vs_theory}
\end{figure*}

For each $n=4, 6, 8, \dots, 28$, we randomly generate $\#_G(n)$ $n$-node 3-regular graphs using the Julia language's Graphs library. For each graph $G$, we randomly generate $\#_b(n)$ LO bitstrings (as described in Section \ref{sec:classicalMaxCut}) to be used as initial cuts for the Warm-Start QAOA algorithm. For each $n$, this yields a total of $\#_G(n) \cdot \#_b(n)$ graph-bitstring $(G,b)$ pairs where $\#_G(n), \#_b(n)$ are functions of $n$ as seen in Equation \ref{eq:gbspairs} below. Due to computational limitations, we consider a smaller number of such pairs for higher $n$.

\begin{equation}
   \begin{tabular}{c}$(G,b)$ pairs \\ on $n$-vertex\\ graphs\end{tabular}: 
\begin{cases}
    \#_G = 30, \#_b = 10, & \text{if } 4 \leq n \leq 18\\
    \#_G = 10, \#_b = 10, & \text{if } 20 \leq n \leq 24\\
    \#_G = 6, \#_b = 3, & \text{if } 26 \leq n \leq 28
\end{cases}
\label{eq:gbspairs}
\end{equation}

For each $(G,b)$ pair, we simulate the Warm-Start QAOA on graph $G$ with warm-start $\ket{b_\theta}$ for all values of tilt angle $\theta = 0^\circ,1^\circ,\dots,90^\circ$ for circuit depth $p=1$; at $n=28$ the tilt angle $\theta$ is instead calculated in $5^\circ$ increments due to computational limitations. The QAOA simulations are performed with JuliQAOA \cite{Golden_2023}, a software library written in Julia for simulating QAOA circuits.

This work considers a variety of parameter optimization schemes and a variety of performance metrics which will be discussed in more detail throughout the remainder of the paper. Unless otherwise stated, all problem sizes $n$ plotted in the figures are averaged over all graph-bitstring pairs in Equation \ref{eq:gbspairs}.

\section{Parameter Optimization}
\label{sec:basin_hopping_optimizaiton}
We first consider Warm-Start QAOA with a simple parameter optimization scheme: the default basin-hopping optimizer in JuliQAOA. In Figures \ref{fig:bh_vs_theory}a and \ref{fig:bh_vs_theory}b, for $n=10,16,22,28$, the average approximation ratios of Warm-Start QAOA (with basin-hopping) are compared against Tate's theoretical bounds (see Section \ref{sec:warmStartQAOA}); the approximation ratios for the simulations are averaged over all graph-bitstring pairs for each $n$.



Figures \ref{fig:bh_vs_theory}a and \ref{fig:bh_vs_theory}b show approximation ratio achieved on the vertical axis for different values of the warm-start tilt angle ($\theta$) on the horizontal axis. Figure \ref{fig:bh_vs_theory}a shows the results of the basin-hopping parameter optimization scheme for $\#_b$ randomly generated bitstrings, which are then locally optimized. Figure \ref{fig:bh_vs_theory}b shows the results for the same basin-hopping parameter optimization scheme, except the $\#_b$ bitstrings are generated using the GW algorithm hyperplane rounding procedure followed by a local search on the bitstrings produced from the GW algorithm (Section II-B). Figure \ref{fig:bh_vs_theory}c and \ref{fig:bh_vs_theory}d show the corresponding scaling of the averaged approximation ratio (y-axis) as the problem size $n\in [4,28]$ (x-axis) increases for several tilt angles ($\theta$).

To understand these plots better we note that, at $\theta=0^\circ$, the results of the single-round Warm-Start QAOA are by definition equivalent to the classically obtained warm-start cut value (since neither phase separator nor mixer have any effect on the state at  $\theta=0^\circ$ as they aim to rotate around the $Z$-axis), whereas the performance at $\theta=90^\circ$ is identical to the single-round standard QAOA with the Pauli-$X$ mixer. Thus, in order for Warm-Start QAOA to outperform both its classical starting state and the standard QAOA variation, these plots would need to show a maximum at some $\theta$ other than at $0^\circ$ or $90^\circ$.

Comparing the approximation ratios for the basin-hopping and Tate’s theoretical lower bounds for  $\theta \in \{0^\circ,90^\circ\}$ in Figures \ref{fig:bh_vs_theory}a and \ref{fig:bh_vs_theory}b, it is seen that the numerical results show, in practice, a strong improvement of this Warm-Start QAOA over Tate's worst case theoretical lower bounds. The upward slope to the peak for $\theta = 60^\circ$, was anticipated from Tate's theoretical work, and can be seen in the orange lower bounds curve, though in a less pronounced fashion. This is due to the fact that for $k$-regular graphs with odd $k$, there exists circuit parameters $\gamma$ and $\beta$ that allow Warm-Start QAOA to exactly recover the initial cut at $\theta = 60^\circ$ \cite{tate2024guaranteeswarmstartedqaoasingleround}.

In Figures \ref{fig:bh_vs_theory}c and \ref{fig:bh_vs_theory}d we can see the basin-hopping parameter optimization scaling of the problem size $n$ compared to averaged approximation ratio for both LO warm-starts and GW LO warm-starts, ranging from $n = 4$ to $n = 28$. The horizontal red-dashed lines show the provable approximation ratio guarantees for Max-Cut on 3-regular graphs for QAOA at depth $p=1$ and the best known algorithm for cubic graphs \cite{halperin2004max}. 
The purple curve shows the scaling of the average approximation achieved with tilt angle $\theta=0^\circ$, i.e. the classically optimized cut. The yellow curve shows tilt angle $\theta=90^\circ$, which, as previously discussed, represents the standard QAOA baseline. The two other curves, blue and green, represent tilt angles $\theta=45^\circ$ and $\theta=60^\circ$. These were chosen as representations for the behavior seen in Figures \ref{fig:bh_vs_theory}a and \ref{fig:bh_vs_theory}b. 

Initially, as expected from Tate's lower bounds, the performance of the Warm-Start QAOA deteriorates. However, as the tilt angle approaches $60^\circ$ the classical state is nearly, or even sometimes recovered with the basin-hopping optimization procedure. The scaling nature of Warm-Start QAOA appears to flatten out for both plots. While further experiments would ideally be conducted for larger problem sizes, we quickly run into computational limits of current HPC systems, particularly considering that we need to average these calculations over many instances. Figures \ref{fig:bh_vs_theory}c and \ref{fig:bh_vs_theory}d seem to suggests some type of asymptotic behavior arising at higher $n$ for this Warm-Start QAOA, which we take as numerical evidence (partiularly for the Goemans-Williamson case in Figure  \ref{fig:bh_vs_theory}d) that we are indeed seeing the asymptotic behavior of Warm-Start QAOA.

\begin{figure}[tbp]
\centerline{\includegraphics[scale=0.38]{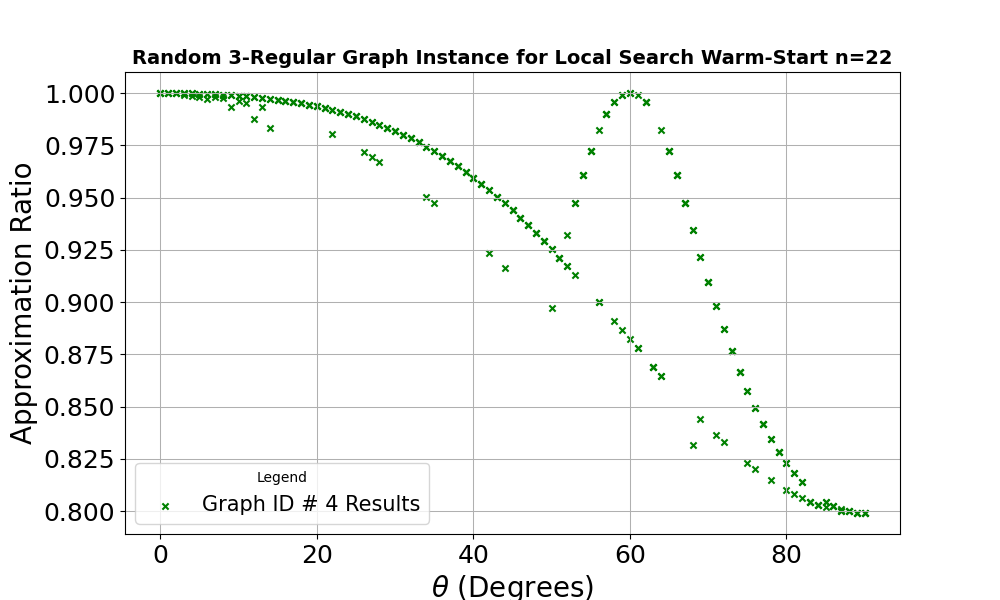}}
\caption{Approximation ratio for a single instance of a randomly generated 3-regular graph, where 3 of the randomly generated bitstrings converge to the same locally optimized bitstring. }
\label{fig:n22pid}
\end{figure}

Overall, the basin-hopping curves in Figures \ref{fig:bh_vs_theory}a and \ref{fig:bh_vs_theory}b appear quite smooth, but it is apparent that the jaggedness implies there are many problem instances where basin-hopping simply fails to find an optimal $\beta$ and $\gamma$. This can be visually observed when looking deeper into a specific problem instance, as seen in Figure \ref{fig:n22pid}. The green curve in Figure \ref{fig:n22pid} plots a randomly generated 3-regular graph instance for $n=22$, where 3 of the randomly generated bitstrings converge to the same LO bitstring used to build the Warm-Start QAOA (the graph instance runs on 3 identical LO bitstrings).

For the same tilt angle $\theta$ with identical LO bitstring for this 3-regular graph instance, the basin-hopping parameter optimization appears to struggle at times to find the optimal values of $\beta$ and $\gamma$ respectively. In some cases it recovers the classical starting state,  while other times it does substantially worse (this basin-hopping behavior is expected and discussed further in Section \ref{sec:improving_optimization}). Furthermore, even when basin-hopping finds what appears to be an optimal set of parameters, it only ever recovers the classical starting cut.  This implies that when using a relatively straight-forward implementation of Warm-Start QAOA, it does not yield approximation ratios that outperform the initial classical cut. In fact, there were no graph-bitstring pairs that the Warm-Start QAOA, with basin-hopping parameter initialization and optimization, for which it found an approximation ratio greater than the classical cut used to build the Warm-Start QAOA.

To avoid this, at times, sporadic difference in approximation ratio, we move our focus to other circuit parameter initialization schemes. In order to improve Warm-Start QAOA performance, we focus on finding better ways to initialize and optimize the $\beta$ and $\gamma$ angles beyond basin-hopping.

As can be seen from the plots in Figure \ref{fig:bh_vs_theory}, both the GW algorithm with LO bitstrings and the LO bitstring warm-starts produce similar behavior. Due to this fact we focus our attention on the randomly generated LO bitstring warm-starts to avoid redundant plots moving forward. Additionally, to maintain consistency in relation to the divergence pronounced in Figure \ref{fig:n22pid}, we maintain a problem size of $n=22$ for fair comparison in Figures \ref{fig:bh_vs_lblm} and \ref{fig:separate_regions}.

\section{Improving Warm-Start Parameter Initialization and Optimization}
\label{sec:improving_optimization}

Figure \ref{fig:bh_vs_theory} shows that a relatively straight-forward implementation of Warm-Start QAOA does not yield approximation ratios that outperform the initial classical cut. In order to improve Warm-Start QAOA performance, we focus on finding better ways to initialize and optimize the $\beta$ and $\gamma$ circuit parameters beyond basin-hopping. Instead of basin-hopping parameter initialization and optimization for the numerical simulations, we use the $\beta$'s and $\gamma$'s from Tate's lower bounds  to initialize the Warm-Start QAOA for each tilt angle $\theta$. 

\begin{figure}[tbp]
\centerline{\includegraphics[scale=0.38]{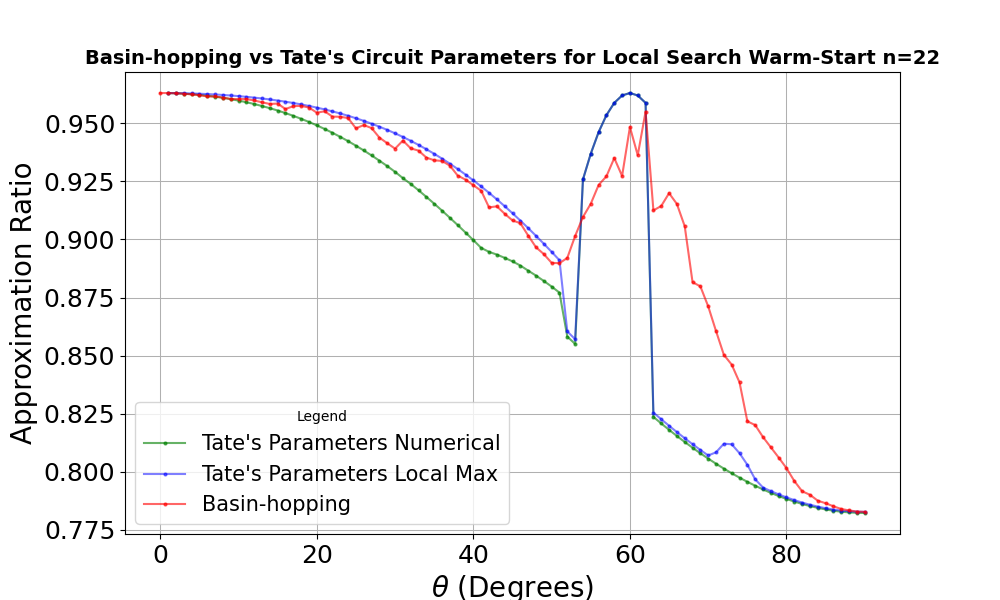}}
\caption{Average approximation ratio achieved with basin-hopping (red curve) vs numerical results for Tate's lower bounds with no further optimizations (green curve), and after local maximum optimization (blue curve) for each tilt angle $\theta$.}
\label{fig:bh_vs_lblm}
\end{figure}

Additionally, we subsequently run a local maximum search of the landscape to see if there exists $\beta$'s and $\gamma$'s that further improve the approximation ratio beyond the initial $\beta$ and $\gamma$, in hopes of obtaining a cut greater than the classical cut used to build the warm-start. Figure \ref{fig:bh_vs_lblm} shows the average approximation ratio achieved for problem size $n=22$ using Tate's theoretical circuit parameters without further optimization (green curve) and with local maximum optimization (blue curve). The red curve is the basin-hopping implementation as described previously in Section \ref{sec:basin_hopping_optimizaiton}.

Figure \ref{fig:bh_vs_lblm} highlights several key observations. First and foremost, it is evident that the blue curve, representing the approximation ratio after conducting a local maximum optimization on Tate's theoretical circuit parameters, is often significantly above the green curve, which corresponds to the averaged approximation ratio using Tate's theoretical circuit parameters without further optimization. This suggests that the local maximum search effectively enhances the performance of the Warm-Start QAOA, yielding better parameter settings than the initial theoretical $\beta$ and $\gamma$ provided by Tate's lower bounds.

The behavior near $\theta = 60^\circ$ is particularly insightful. The blue curve remains above the red curve and appears far smoother, further indicating, as mentioned previously, that basin-hopping struggles at times to find the optimal angles. This is evidenced by the red line's performance, which occasionally lags behind, particularly around $\theta = 60^\circ$ for reason observed in  Figure \ref{fig:n22pid} in the previous section. Despite this improvement, the fact that the red basin-hopping curve sometimes exceeds the blue line, especially for $\theta > 63^\circ$, indicates that even with the enhanced scheme, we are not consistently finding the optimal angles. This observation is corroborated by the discontinuities in the blue line around $\theta = 52^\circ$ and $\theta = 63^\circ$, suggesting that the current optimization approach may be insufficient.

These findings lead us to conclude that while initializing the Warm-Start QAOA with Tate's theoretical circuit parameters, and additionally using local maximum optimization, provides a noticeable improvement over the basin-hopping method on average, it is not a foolproof solution. The inconsistencies observed across different tilt angle $\theta$ regimes suggests that a more sophisticated and nuanced initialization and optimization scheme is necessary.

\begin{figure*}[t]
  \centering
  \subfloat[Average approximation ratio achieved with circuit parameter \textbf{region 1}: $\beta = 0.486 $ $\gamma  = 0.5634$.]{%
    \includegraphics[width=0.49\textwidth]{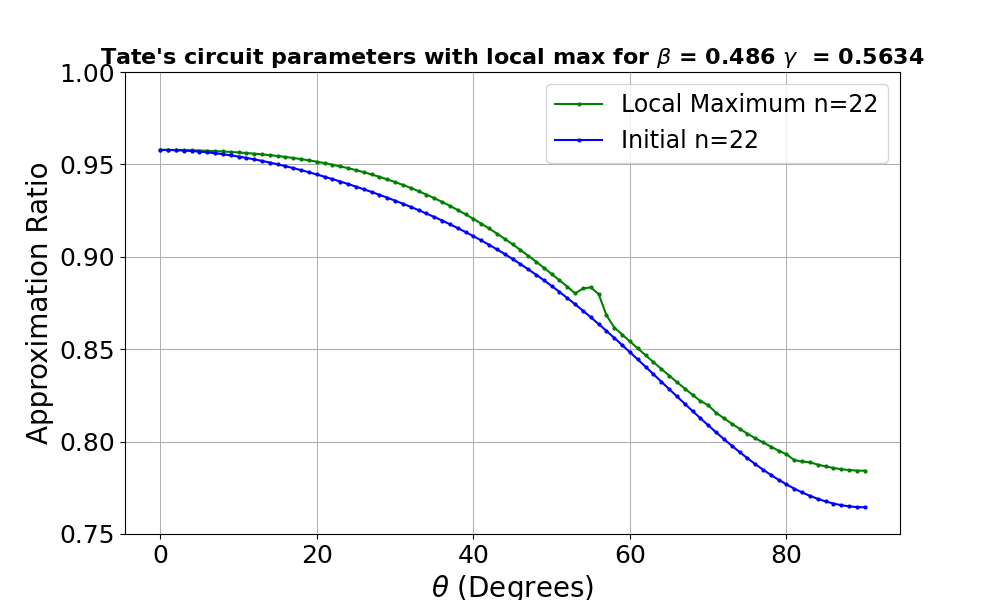} 
  }%
  \hfill
  \subfloat[Average approximation ratio achieved with circuit parameter \textbf{region 2}: $\beta = 1.5705 $ $\gamma  = 3.142$.]{%
    \includegraphics[width=0.49\textwidth]{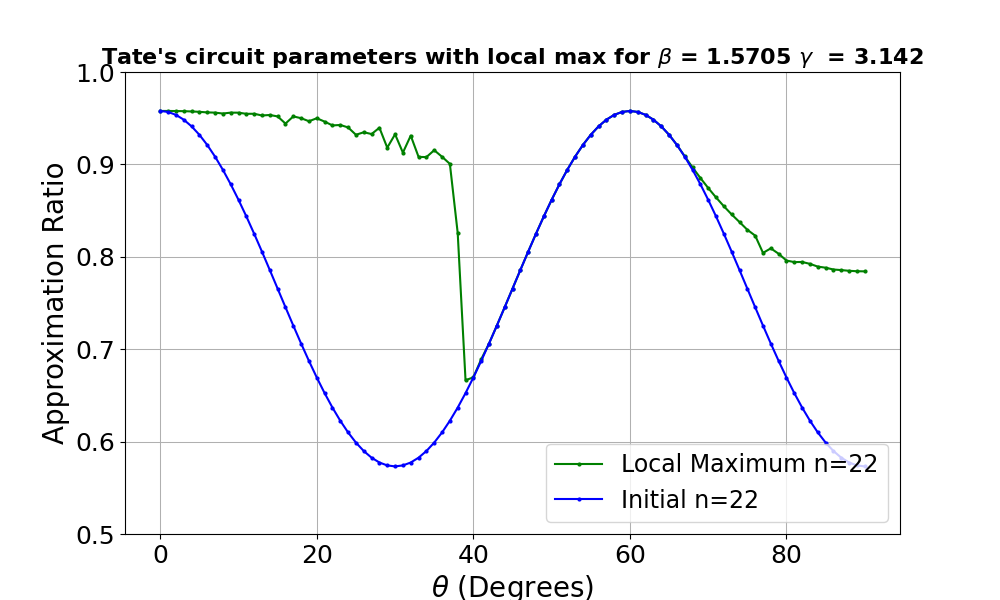}%
  }\\
  \vspace{0.5cm} 
  \subfloat[Average approximation ratio achieved with circuit parameter \textbf{region 3}: $\beta = 0.3996 $ $\gamma  = 2.5144$.]{%
    \includegraphics[width=0.49\textwidth]{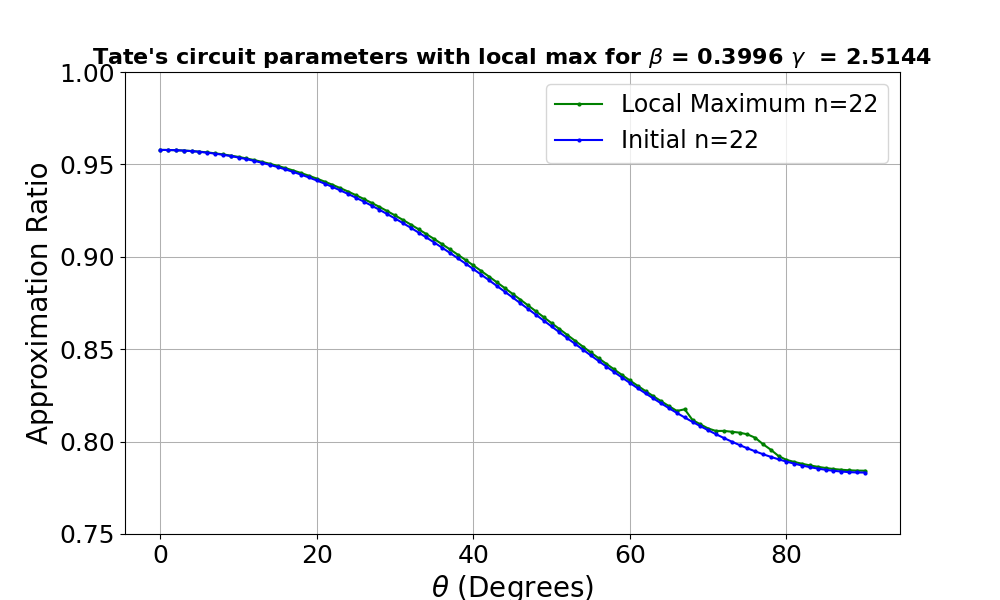}%
  }%
  \hfill
  \subfloat[Average approximation ratio achieved with basin-hopping (red curve), combined Tate's parameter regions without optimization (blue curve) and after local maximum optimization (green curve). The two regions that dominate after local maximum optimization (green curve) are angle region 1 from $1 \leq \theta \leq 51$ and angle region 2 with $52 \leq \theta \leq 89$. Angle region 3 does not converge to a better local maximum than either regions 1 or 2. ]{%
    \includegraphics[width=0.49\textwidth]{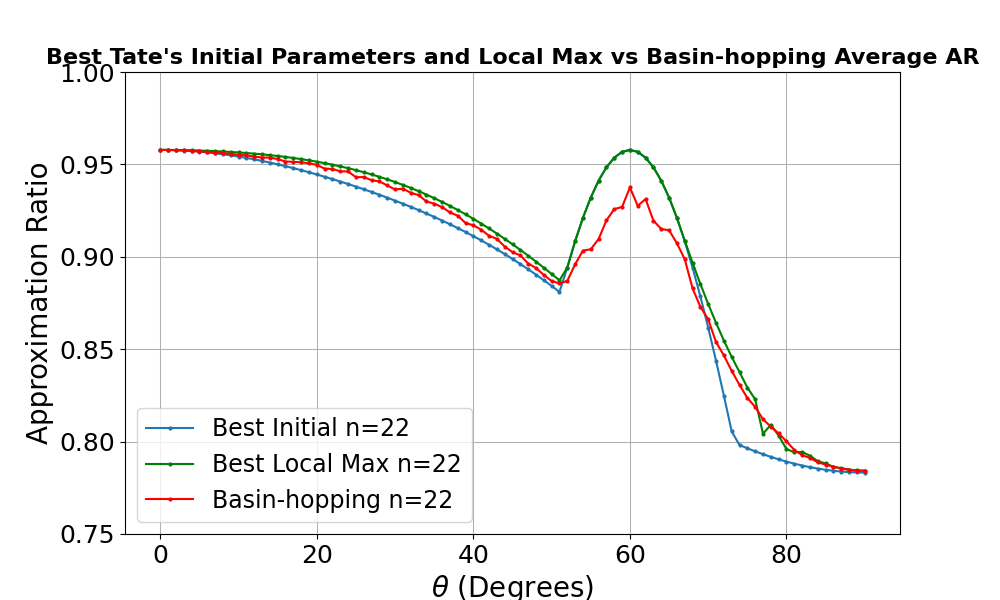}%
  }
  
  \caption{}
  \label{fig:separate_regions}
\end{figure*}
Tate et al. \cite{tate2024guaranteeswarmstartedqaoasingleround} observed that for varying values of tilt-angle $\theta$, that the theoretical optimal circuit parameters (corresponding to their theoretical lower bounds) were clustered into three groups (see Figure 17 of \cite{tate2024guaranteeswarmstartedqaoasingleround} with $\kappa=2/3$). To further improve upon the discontinuities in Figure \ref{fig:bh_vs_lblm} we consider three $(\gamma,\beta)$ pairs that are representative of the three regions described by Tate et. al \cite{tate2024guaranteeswarmstartedqaoasingleround}. We then run numerical simulations with $\beta$'s and $\gamma$'s from each of these three regions; we present the results for $n=22$ in Figures \ref{fig:separate_regions}a, \ref{fig:separate_regions}b, and \ref{fig:separate_regions}c. The blue curve shown is the average approximation ratio achieved with Tate's initial circuit parameters $\beta$ and $\gamma$, prior to a local maximum optimization, shown by the green curve. We then draw upon the best average approximation values from the three separate regions and combine them further into Figure \ref{fig:separate_regions}d, plotted against the average approximation ratio achieved with the basin-hopping technique in red.

When comparing the plots in Figure \ref{fig:separate_regions} to the graph instance with multiple identical LO bitstrings in Figure \ref{fig:n22pid}, the sporadic behavior associated with basin-hopping becomes more understandable. It appears the discrepancy in approximation ratio can be explained by the basin-hopping scheme finding a local optimum associated with one of Tate's 3 circuit parameter regions. At times the $\beta$ and $\gamma$ found by basin-hopping finds expectation values associated the circuit parameters in region 1, while at other times it finds expectation values associated with circuit parameter regions 2 or 3. Moreover, as seen in Figures \ref{fig:separate_regions}a, \ref{fig:separate_regions}b, and \ref{fig:separate_regions}c, these regions themselves can be optimized further, thus leading to the inconsistent performance of the basin-hopping technique observed in Figure \ref{fig:n22pid}.

The technique of selecting a $\beta$ and $\gamma$ from each of the 3 circuit parameter regions found by Tate et al. \cite{tate2024guaranteeswarmstartedqaoasingleround} produces fluid curves and shows the importance of theory as a foundation for experimental simulations in Warm-Start QAOA research. In addition, it is observed that local maximum optimization techniques are extremely useful in improving performance of the Warm-Start QAOA circuit parameters $\beta$ and $\gamma$. The green curve in Figure \ref{fig:separate_regions} shows an overall improvement over basin-hopping across the approximation ratio averages. It also appears that the optimal circuit parameters are found, however, for all graph-bitstring pairs, this warm-start still fails to produce a higher approximation ratio greater than that of the classical starting state. The classical cut is recovered, but a better solution is never found.

\section{A Novel Metric: Optimizing for Better Solution Probabilities}
\label{sec:novel_metric}
Our results thus far paint a bleak picture for single-round Warm-Start QAOA: the classical starting cut is better than or equal to the the expected cut size found. While the expected approximation ratio is the standard metric in the QAOA literature, the probability of sampling an optimum solution, often referred to as ground state probability (GSP), is sometimes used. Appendix \ref{subsec:GSP} Figure \ref{fig:aligned_gs} shows the ground state sampling probabilities achieved with basin-hopping optimized for the standard expectation value. While we see similar behavior to that of the approximation ratio, we note that Warm-Start QAOA does not outperform the GSP of the initial classical cut (i.e., the value at $\theta=0$ or the likeliehood that the optimium solution is found by the classical algorithm) for any other tilt angle $\theta$. Additionally, we remark that, in practice, optimizing the circuit parameters for GSP cannot be realistically done since optimizing for GSP requires knowledge of the optimal solution a priori.
 
Due to this fact of GSP requiring knowledge a priori of the optimal solution, we consider other metrics by which to analyze the Warm-Start QAOA. In order to change our frame of reference, we look further into the statevector probabilities for cuts whose value is greater than the value of the initial cut used to build the Warm-Start QAOA. We define Better Solution Probability (BSP) formally in Equations \ref{bsp_1} and \ref{bsp_2}

If 
\begin{equation}
    \ket{\psi} = \sum_{x \in \{0,1\}^n} c_x\ket{x}
    \label{bsp_1}
\end{equation}
is the output of Warm-Start QAOA with initial solution $b$,
then,
\begin{equation}
    \text{BSP} = \sum_{\substack{x \in \{0,1\}^n:\\ \cut(x) > \cut(b)}} |c_x|^2.
    \label{bsp_2}
\end{equation} 

\begin{figure*}[t]
  \centering
  \subfloat[Averaged Better Solution Probability (BSP) vs $\theta$ achieved using Tate's circuit parameters (CP) from region 1.]{%
    \includegraphics[width=0.49\textwidth]{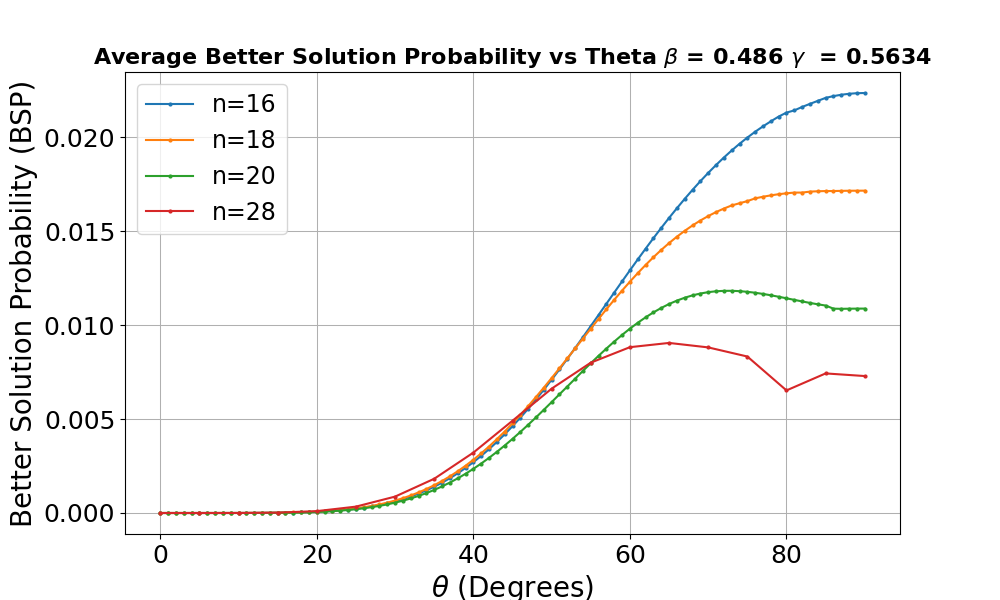} 
  }%
  \hfill
  \subfloat[Averaged Better Solution Probability (BSP) vs $\theta$ using Tate's circuit parameters (CP) from region 1, basin-hopping (\textbf{optimized for BSP}), and grid search before and after local max (\textbf{optimized for BSP}).]{%
    \includegraphics[width=0.49\textwidth]{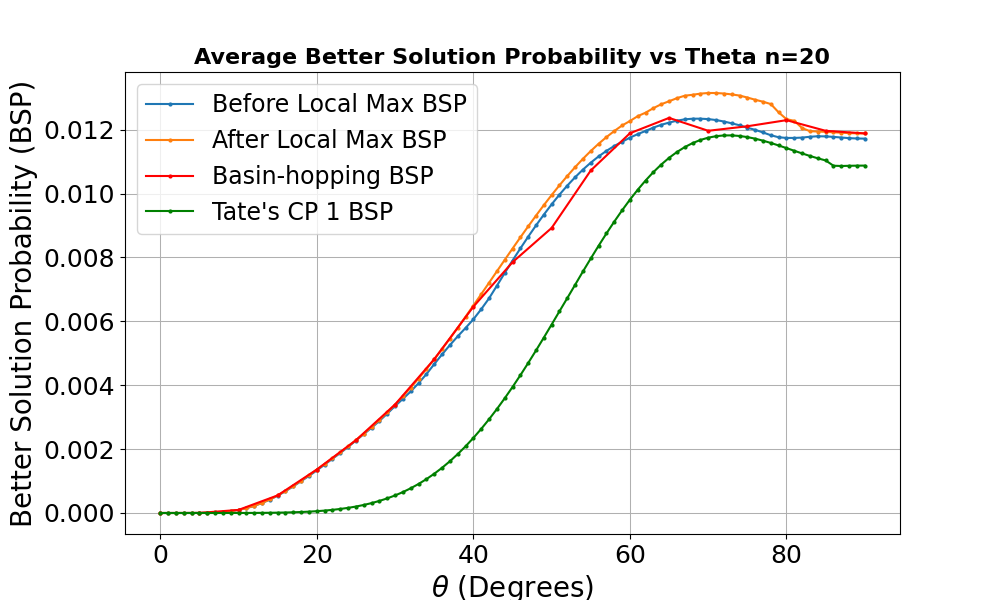}%
  }
    \hfill
  \subfloat[Warm-start basin-hopping optimization for BSP (red curve) versus warm-start basin-hopping optimization for expectation value (green curve) vs $\theta$]{%
    \includegraphics[width=0.49\textwidth]{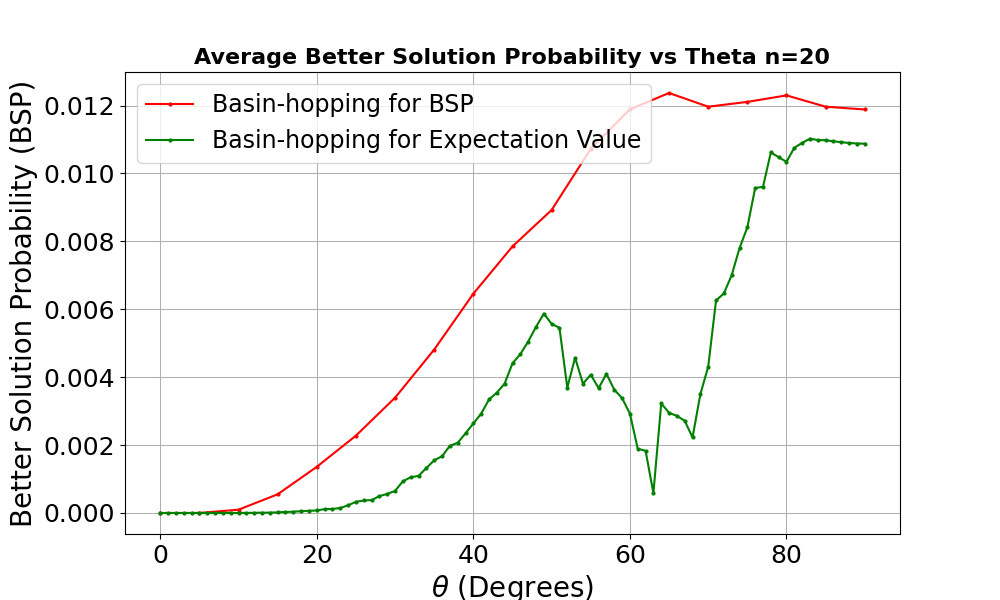}%
  }
  \caption{}
  \label{fig:bsp}
\end{figure*}

Our BSP metric bears some resemblance to the concepts of Threshold QAOA \cite{Golden_2021}, which considers solutions with approximation ratios above a certain threshold. There are, however, noticeable differences. First, in the context of warm-starts, the value to beat for BSP is already determined by the warm-start; whereas for Threshold-based QAOA, the threshold itself is a hyper-parameter that needs to be decided. Secondly, while Threshold QAOA gives consideration to cuts or solutions above some threshold, their approach is vastly different, e.g. they optimize circuit parameters for expectation value (i.e. approximation ratio) and approximation ratio is the final "figure of merit" that is considered. In contrast, in this paper, BSP \emph{\textbf{is}} the parameter optimization objective.

Figure \ref{fig:bsp}a, compares the averaged BSP for problem sizes of $n=16$ to $n=28$. We find for Tates’s $\beta = 0.486$ and $\gamma = 0.5634 $ region 1 circuit parameters, that there are instances where there exists non-trivial tilt angles $\theta$ at which the Warm-Start QAOA algorithm can sample larger cuts or better solutions with a higher probability on average than standard QAOA at $\theta = 90^\circ$. We note that at $\theta = 0^\circ$ BSP will be $0.0$ for reasons previously mentioned in section \ref{sec:basin_hopping_optimizaiton}. As can be seen in Figure \ref{fig:bsp}a, the BSP advantage of Warm-Start QAOA appears to become more pronounced as problem size $n$ increases. The most drastic BSP advantages for the Warm-Start QAOA are seen for $n\geq 20$. The peak for $n = 20$ occurs at the tilt angle $\theta = 75^\circ$ and at $\theta = 65^\circ$ for $n=28$ respectively. It is important to state that while there are many graph-bitstring pairs where the Warm-Start QAOA approach has a higher BSP for problem sizes $n\leq 18$ than standard QAOA ($\theta=90^\circ$), this was not the average case until higher $n$ as it appeared for $n\geq 20$.

The results and insight observed from Figure \ref{fig:bsp}a, inspires a shift in our optimization approach with respect to BSP as a metric. Rather than using the standard QAOA objective function, which maximizes circuit parameters $\beta$ and $\gamma$ for expectation value, we instead create a new objective function to maximize the circuit parameters for BSP. By doing so, we aim to exploit the unique strengths of Warm-Start QAOA, where the probability of finding a better solution than the initial classical solution can be enhanced. We note that the circuit parameter regions found by Tate et al. \cite{tate2024guaranteeswarmstartedqaoasingleround} were discovered under the assumption of an objective function optimizing for expectation value.

Figure \ref{fig:bsp}b shows the averaged BSP achieved for problem size $n=20$ with an objective function optimizing for BSP, using several different optimization techniques previously discussed in this paper. Due to the novel nature of the metric, basin-hopping required substantially more steps to find greater values of BSP than when optimized for expectation value. For these reasons $n=20$ was the largest problem size for which we were able to receive data under computational constraints. The green curve represents the average BSP achieved only using $\beta$ and $\gamma$ from Tate's region 1 circuit parameters. The red curve represents the average BSP achieved using the basin-hopping parameter initialization and optimization scheme. The blue and orange curve show the average BSP achieved using a 40x40 grid search over $\beta$ and $\gamma$.

A first observation from the curves in Figure \ref{fig:bsp}b is that all three optimization techniques, basin-hopping, grid search, and grid search with local max, improve the overall average of BSP, suggesting that BSP, as a novel metric, can successfully be optimized for with respect to circuit parameters $\beta$ and $\gamma$. Secondly, it appears that grid search with local max provides the greatest advantage over the other two optimization schemes, improving upon the best BSP acheived with Tate's region 1 circuit parameters by more than 10\% from $0.0118$ to $0.0131$. Additionally, the Warm-Start QAOA BSP advantage is maintained as seen from the peak reached at $\theta = 70^\circ$ for the orange curve. 

Additionaly, Figure \ref{fig:bsp}c shows the differences achieved when basin-hopping is used as a parameter optimization technique for BSP (red curve) versus for expectation value (green curve). As can be observed from the plot, using basin-hopping optimized for BSP shows improvement over the BSP achieved when using the standard expectation value objective.

Overall, we observe improvement for BSP sampling capabilities for standard QAOA ($\theta=90^\circ$), indicating its potential usefulness as a generalized metric in the field of QAOA research. While expectation value remains a general benchmark for the performance of QAOA, optimizing for our BSP metric provides a potentially more powerful way of assessing and improving the performance of Warm-Start QAOA, offering a new direction for achieving quantum advantage in combinatorial optimization problems.

\section{Discussion and Conclusion}
The results from our experimental study on Warm-Start QAOA reveal a complex landscape of algorithmic performance that both confirms and challenges existing theoretical insights. As anticipated, our empirical findings demonstrate that Warm-Start QAOA, even with a simple basin-hopping parameter optimization scheme, significantly outperforms the theoretical worst-case lower bounds established by Tate et al. for 3-regular Max-Cut instances. This gap between theoretical expectations and practical outcomes underscores the importance of empirical validation in the study of quantum algorithms.

However, our investigation also uncovered some limitations of the Warm-Start QAOA approach, particularly at the single-round depth ($p=1$). Despite the promise of warm-starting from a classically optimized bitstring, our results consistently showed that the performance of the Warm-Start QAOA did not surpass the classical starting cut for any choice of the tilt angle $\theta$. This suggests that while Warm-Start QAOA can recover the initial classical solution, it does not improve upon it within the constraints of a single round, which raises questions about its practical advantage in a scenario where we are limited to $p=1$.

To address this limitation, we introduced a novel figure-of-merit, the Better Solution Probability (BSP), which measures the probability that Warm-Start QAOA identifies a better solution than the initial classical cut. Our results indicate that, under this metric, Warm-Start QAOA frequently performs best at non-trivial tilt angles (i.e., $0^\circ < \theta < 90^\circ$), suggesting that there is still untapped potential in this algorithmic variant. Moreover, the BSP metric proves to be more practical for parameter optimization than other metrics like ground state probability, as it does not require prior knowledge of the optimal solution, rather just sampling of the state vector.

\subsection{Future Work}
Our study opens up several avenues for future research. One of the most immediate directions is to extend the Warm-Start QAOA to higher circuit depths ($p > 1$). Our current work focuses on single-round QAOA for which Tate et. al had proven theoretical lower bounds. As additional rounds of Warm-Start QAOA are added, the increased circuit depth greatly complicates the optimization landscape for $\beta$ and $\gamma$, making analytical derivations for theoretical lower bound tilt angles $\theta$ extremely challenging. However, increasing the circuit depth could potentially unlock more of the algorithm's theoretical advantages, allowing it to surpass the classical starting solution and standard QAOA more consistently.

Additionally another direction we would like to explore is expanding the scope of our experiments beyond the 3-regular Max-Cut problem. Warm-Start QAOA and BSP can be applied to a broader set of combinatorial optimization problems, where the initial state is derived from different classical algorithms tailored to those specific problems. Furthermore, exploring other graph families, beyond 3-regular graphs, could provide deeper insights into the algorithm's generalizability and robustness. These future investigations will be essential for determining the full potential and practical applicability of Warm-Start QAOA and BSP in solving real-world optimization problems.


\section{Acknowledgments}
\label{section:acknowledgments}
This work was supported by the U.S. Department of Energy through the Los Alamos National Laboratory. Los Alamos National Laboratory is operated by Triad National Security, LLC, for the National Nuclear Security Administration of U.S. Department of Energy (Contract No. 89233218CNA000001). The research presented in this article was supported by the Laboratory Directed Research and Development program of Los Alamos National Laboratory under project number 20230049DR as well as by the NNSA's Advanced Simulation and Computing Beyond Moore's Law Program at Los Alamos National Laboratory.  

\section{Appendix}

\subsection{Ground State Probabilities}
\label{subsec:GSP}
\begin{figure}[h]
\centerline{\includegraphics[scale=0.38]{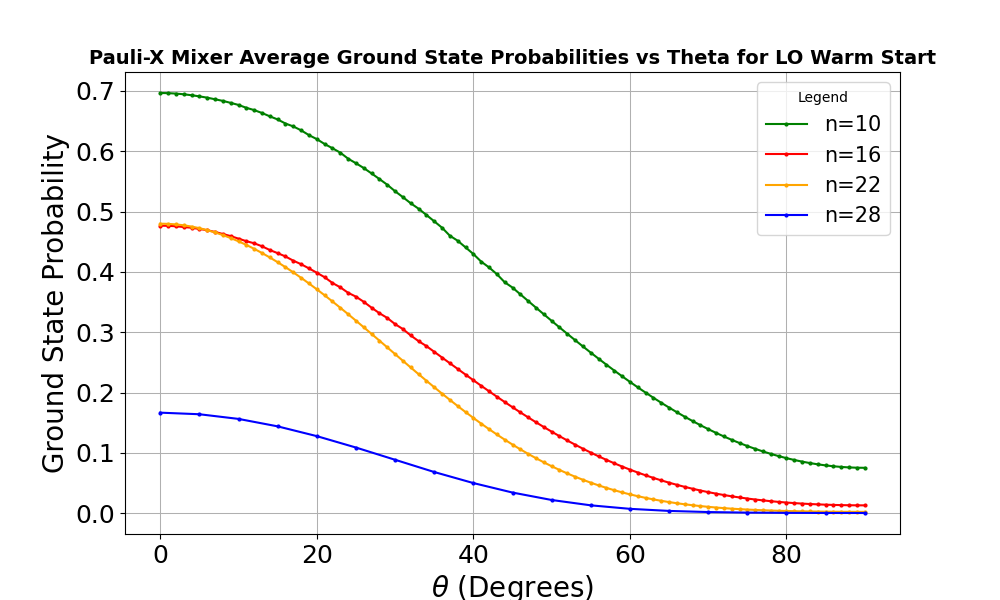}}
\caption{Pauli-$X$ mixer average ground state probabilities achieved with LO warm-start and basin-hopping (optimizing for expectation value).}
\label{fig:x_gs}
\end{figure}

\begin{figure}[htbp]
\centerline{\includegraphics[scale=0.38]{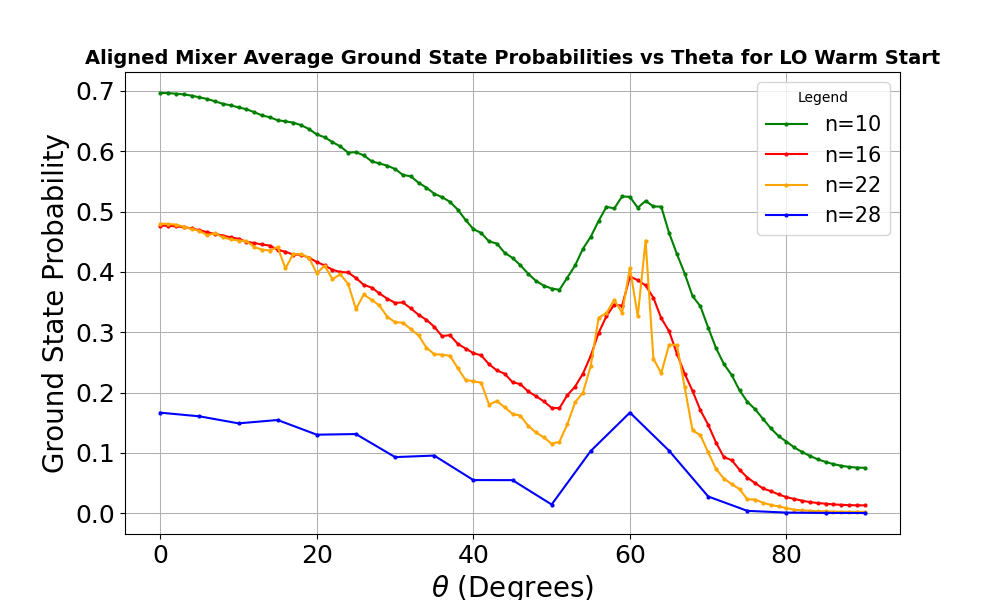}}
\caption{Aligned mixer average ground state probabilities achieved with LO warm-start and basin-hopping (optimizing for expectation value).}
\label{fig:aligned_gs}
\end{figure}

Ground state probability, GSP, is sometimes used as a metric to judge the performance of QAOA. Figures \ref{fig:x_gs} and \ref{fig:aligned_gs} show the average GSP achieved for both the standard Pauli-$X$ mixer and the aligned mixer considered in this paper for problem sizes $n=10$ to $n=28$. As can be seen from the plots, there is a distinct behavior difference between the Pauli-$X$ mixer and the aligned Mixer where the GSP constantly decreases for the X mixer, but spikes, as anticipated for the Aligned Mixer at $\theta=60^\circ$. We reiterate, however, that in practice, optimizing the circuit parameters for GSP cannot be realistically done since optimizing for GSP requires knowledge of the optimal solution a priori.

\subsection{Aligned Mixer vs Pauli-$X$ Mixer Approximation Ratios}
\begin{figure}[htbp]
\centerline{\includegraphics[scale=0.38]{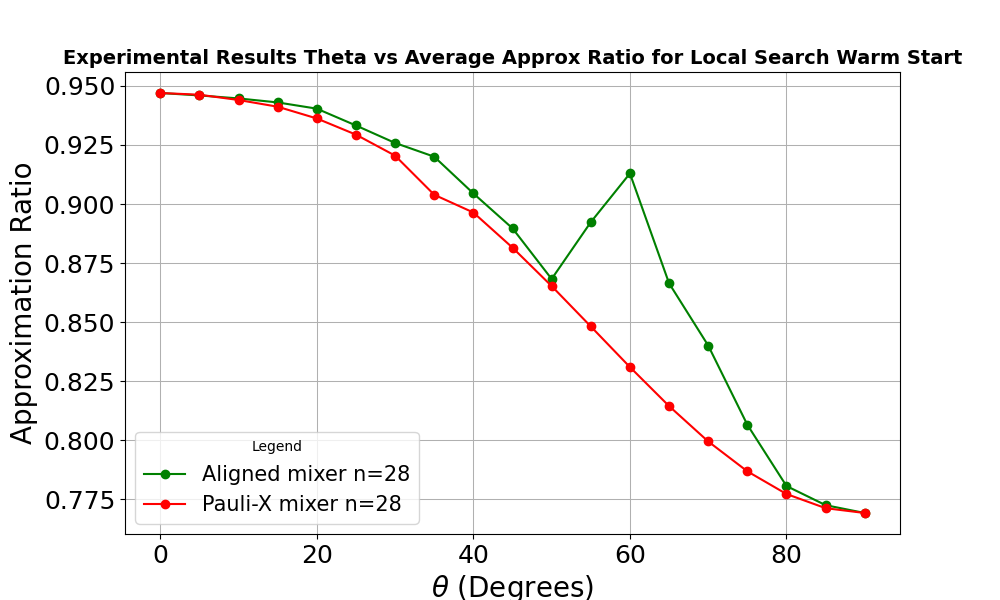}}
\caption{Pauli-$X$ mixer vs Aligned mixer for $n=28$ average approximation ratio achieved with basin-hopping (optimizing for expectation value).}
\label{fig:aligned_vs_x}
\end{figure}

\begin{figure}[htbp]
\centerline{\includegraphics[scale=0.38]{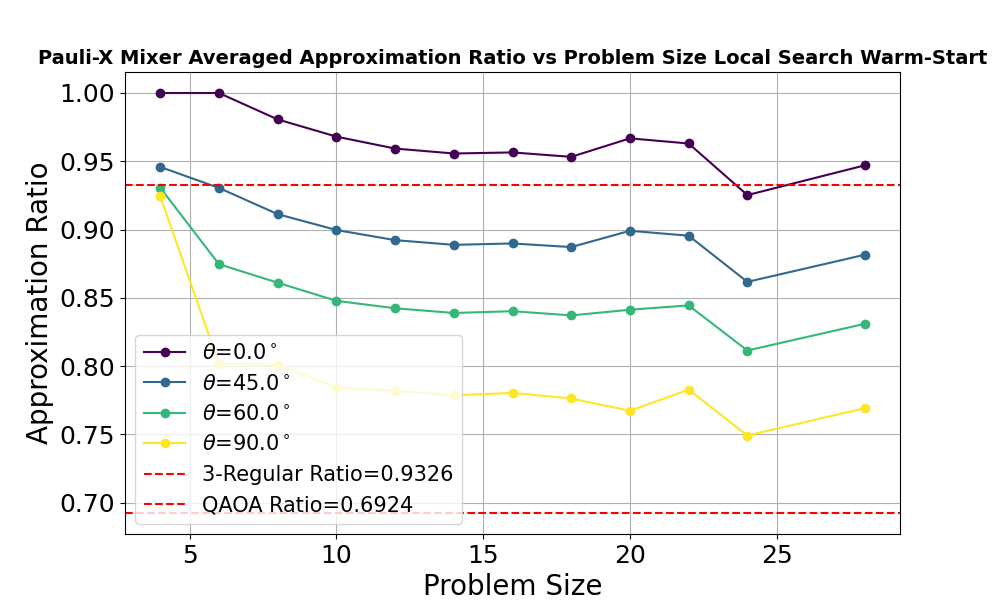}}
\caption{Pauli-$X$ mixer average approximation ratio vs problem size $n$ basin-hopping circuit parameter optimization with \textbf{Local Search} warm-start state for $\theta = 0^\circ , 45^\circ, 60^\circ, 90^\circ$ tilt angles. Red dashed lines represent the best known 3-Regular Max-Cut approximation ratio \cite{halperin2004max} and Farhi's QAOA 3-Regular Max-Cut approximation ratio guarantee for circuit depth $p=1$. }
\label{fig:x_n_vs_ar}
\end{figure}

The differences in numerical results for Warm-Start QAOA with the aligned mixer and the Pauli-$X$ mixer are presented in Figures \ref{fig:aligned_vs_x} and \ref{fig:x_n_vs_ar}. Figure \ref{fig:aligned_vs_x} shows the average approximation ratio achieved with the basin-hopping parameter optimization and initialization scheme for $n=28$. As seen similar to the differences discussed in Appendix \ref{subsec:GSP}, the Pauli-$X$ mixer continually decreases to $\theta=90^\circ$, while the aligned mixer sees the signature spike at $\theta=60^\circ$.

Figure \ref{fig:x_n_vs_ar} shows the Pauli-$X$ mixer average approximation ratio achieved (y-axis), vs the problem size $n$ (x-axis). Similar to the plots in Section \ref{sec:basin_hopping_optimizaiton}, in Figure \ref{fig:bh_vs_theory}, we plot the average approximation ratios for several tilt angles $\theta$.

\bibliographystyle{IEEEtran}
\bibliography{sources}
\vspace{12pt}

\end{document}